\documentclass[epj]{svjour}

\usepackage{graphics}
\usepackage[latin1]{inputenc}
\usepackage{amsmath}
\begin{document}

\title{
Negative heat capacity for hot nuclei using
formulation\\ from the microcanonical ensemble\\
INDRA Collaboration}

\headnote{Letter to the Editor}

\author{
B.~Borderie\inst{1}
\and S. Piantelli\inst{2}
\and E.~Bonnet\inst{3}
\and R.~Bougault\inst{4}
\and A.~Chbihi\inst{5}
\and J.E.~Ducret\inst{5}
\and J.D.~Frankland\inst{5}
\and E.~Galichet\inst{1,6}
\and D.~Gruyer\inst{4}
\and M.~Henri\inst{5}
\and M.~La~Commara\inst{7}
\and N.~Le Neindre\inst{4}
\and I.~Lombardo\inst{8}
\and O.~Lopez\inst{4}
\and L.~Manduci\inst{4,9}
\and M.~P\^arlog\inst{4,10}
\and R.~Roy\inst{11}
\and G.~Verde\inst{8}
\and M.~Vigilante\inst{7}
}

%
%
\institute{
Institut de Physique Nucl\'eaire, CNRS/IN2P3,
Univ. Paris-Sud, Universit\'e Paris-Saclay, Orsay, France
\mail{borderie@ipno.in2p3.fr}
\and INFN, Sezione di Firenze, Sesto Fiorentino, Italy
\and SUBATECH, IMT Atlantique, Universit\'e de
Nantes, CNRS-IN2P3, Nantes, France
\and Normandie Univ., ENSICAEN, UNICAEN, CNRS/IN2P3, LPC Caen, Caen, France
\and GANIL, (CEA/DRF-CNRS/IN2P3), Caen, France
\and Conservatoire National des Arts et M\'etiers, Paris, France
\and Dipartimento di Fisica `E. Pancini' and Sezione INFN,
Universit\`a di  Napoli ``Federico~II'', Napoli, Italy
\and INFN, Sezione di Catania, Catania, Italy
\and Ecole des Applications Militaires de l'Energie Atomique,
Cherbourg, France
\and Hulubei National Institute for R \& D in Physics and
Nuclear Engineering (IFIN-HH), Bucharest-M\'agurele, Romania
\and Universit\'e Laval, Qu\'ebec, Canada
}
%
%
\abstract{
By using freeze-out properties of multifragmenting hot nuclei produced
in quasifusion central $^{129}$Xe+$^{nat}$Sn collisions at different 
beam energies 
(32, 39, 45 and 50 AMeV) which were estimated by means of a
simulation based on experimental data collected by the $4\pi$    
INDRA multidetector, heat capacity in the thermal excitation energy
range 4 - 12.5 AMeV
was calculated from total kinetic energies and multiplicities at
freeze-out. The microcanonical formulation was employed.
Negative heat capacity which signs a first order phase transition
for finite systems is observed and confirms previous results using
a different method.
%
} 


\maketitle


An important challenge of heavy-ion collisions at intermediate
energies is the identification and characterization of the nuclear
liquid-gas phase transition in hot nuclei, which has been theoretically
predicted for nuclear
matter~\cite{Sch82,Cur83,Jaq83,Mul95,Wel14}. At present one can say that
huge progress has been made even if some points can be deeper
investigated~\cite{WCI06,Bor08,Bor19,Wad19,Lin19}. 
Statistical mechanics for finite systems appeared as a key issue to
progress, revealing specific first-order phase transition signatures
related to the consequences of the local convexity of the 
entropy~\cite{Lab90,Wal94,Gro97,Gulm02,Cho04,Tou09,Cam09}. 
By considering the microcanonical ensemble with energy as extensive
variable, the convex intruder implies a backbending in the temperature
(first derivative of entropy) at constant pressure~\cite{Cho00}
and correlatively a 
negative branch for the heat capacity (second derivative of entropy).
Experimentally, these two converging signatures have been observed
in hot nuclei from different 
analyses~\cite{I79-Bor13,MDA00,NicPHD,I46-Bor02,I63-NLN07}
of homogeneous event samples. It is important to recall here that
signals of phase transition in hot nuclei are only meaningful at
the level of statistical ensembles constructed from the outcome of
carefully selected collisions~\cite{Bor19}.
Another consequence of the entropy curvature anomaly manifests itself
when systems are treated in the canonical ensemble. 
In this case a direct phase transition
signature is the presence of a bimodal distribution of an order
parameter~\cite{Cho01} such as 
the charge (size) of the largest fragment (Z$_{max}$) of multifragmentation
partitions~\cite{I72-Bon09}. 
As far as microcanonical heat capacity calculation is concerned, a
question still subsists. It concerns one hypothesis made in the method
used up to now: the same temperature was associated with both
internal excitation and thermal motion of emitted fragments, which is
not physically obvious if one remembers that the level density is
expected to vanish at high excitation 
energies~\cite{Tub79,Dean85,Koo87,Sou15}.
Moreover it was shown in simulations compared to data 
that widths of the fragment velocity spectra cannot be reproduced 
with the hypothesis of a common temperature for intrinsic and kinetic
degrees of freedom~\cite{I66-Pia08}. 

In the present letter we will give heat capacity information
for experimental data from microcanonical formulae and by ignoring
the hypothesis of a single temperature, which requires the
introduction of a limiting temperature for fragments.

We first briefly recall the method proposed for
measuring heat capacity using partial energy fluctuations
with a microcanonical sampling~\cite{Cho99,Gul00,Cho00}.
The prescription is based on the fact that,
for a given total energy, the
average partial energy stored in a part of the system is a good 
microcanonical thermometer, while the associated fluctuations can be
used to construct the heat capacity.
From experiments the most simple decomposition of the thermal
excitation energy 
is in a kinetic part, $E_{k}$, and a potential part, $E_{pot}$ 
(Coulomb energy + total mass excess).
These quantities have to be determined at freeze-out  and
consequently it is necessary to trace back this
configuration on an event by event basis.
The true configuration needs the knowledge of the freeze-out
volume and of all the particles evaporated from primary hot
fragments including the (undetected) neutrons. Consequently
some working hypotheses are used, constrained by
specific experimental results (see for example~\cite{MDA02}).
Then, the experimental correlation between the kinetic energy per nucleon
$E_{k}$/$A$ and
the thermal excitation energy per nucleon $E^{*}$/$A$ of the considered
system can be obtained event by event 
as well as the variance of the kinetic energy
$\sigma_{k}^{2}$. Note that $E_{k}$ is calculated by subtracting 
the potential part $E_{pot}$ from the thermal excitation energy $E^{*}$ 
and consequently
kinetic energy fluctuations at freeze-out reflect the configurational 
energy fluctuations.
In the microcanonical ensemble with total energy $E^{*}$ the total
degeneracy factor is simply given by the folding product of the
individual degeneracy factors $W_{k} = exp(S_{k}(E_{k}))$ and 
$W_{pot} = exp(S_{pot}(E_{pot}))$ where $S_{k}$($S_{pot}$) is the
entropy of the kinetic (potential) part. 
One can then define for the total
system as well as for the two subsystems the microcanonical temperatures
and the associated heat capacities $C_k$ and $C_{pot}$.
If we consider now the kinetic energy distribution when the total
energy is $E^{*}$ we get
\begin{equation}
\noindent P_{E^*}(E_{k}) = exp(S_{k}(E_{k}) + S_{pot}(E^* -E_{k})).
\label{eq:Kindistri}
\end{equation}
Then the most probable kinetic energy $\overline{E_{k}}$ is defined by
the equality of the partial microcanonical temperatures
$T_{k}(\overline{E_{k}})$ = $T_{pot}(E^{*} - \overline{E_{k}})$ and  
$\overline{E_{k}}$ can be used as the microcanonical thermometer.
An estimator of the
microcanonical temperature of the system can be obtained by inverting the
kinetic equation of state:
\begin{equation}
\noindent < E_{k} > = \langle \sum_{i=1}^{M} a_i \rangle T^2 +
\langle \frac{3}{2} (M-1) \rangle T 
\label{eq:Keos}
\end{equation}
The brackets $\langle\rangle$
indicate the average on events with the same $E^{*}$,
$a_i$ is the level density parameter and M the multiplicity at
freeze-out. \textit{In this expression the same
temperature is associated with both internal excitation and thermal
motion of fragments.}
An estimate of the total microcanonical heat capacity 
is extracted using three equations.
\begin{equation}
\noindent C_k = \frac{\delta <E_k / A >}{\delta T},
\label{eq:CnegM1}
\end{equation}
is obtained by taking the derivative of $<E_k / A >$ with 
respect to $T$.

Using a Gaussian approximation for $P_{E^*}(E_{k})$ the kinetic
energy variance can be calculated as\\
\begin{equation}
A\sigma_{k}^{2} \simeq T^2\frac{C_kC_{pot}}{C_k+C_{pot}};
\label{eq:CnegM2}
\end{equation}
Eq.~(\ref{eq:CnegM2}) can be inverted to extract, from the observed
fluctuations, an estimate of the microcanonical heat capacity:
\begin{equation}
\noindent (\frac{C}{A})_{micro} \simeq C_k+C_{pot} \simeq \frac{C_k^2}
 {C_k - \frac{A \sigma_k^2}{T^2}}.
\label{eq:CnegM3}
\end{equation} 
From Eq.~(\ref{eq:CnegM3}) we see that the specific microcanonical
heat capacity $(C/A)_{micro}$ becomes negative if the normalized
kinetic energy fluctuations $A \sigma_k^{2}/T^2$ overcome $C_k$.
It is interesting to note that the constraint of energy conservation
leads in the phase transition region to larger fluctuations than in
the canonical case where the total energy is free to fluctuate. This
is because the kinetic energy part is forced to share the total
available energy with the potential part: when the potential part
presents a negative heat capacity the jump from ``liquid'' to ``gas''
induces strong fluctuations in the energy partitioning.

Direct formulae have been proposed in Ref.~\cite{Radut02}
to calculate heat capacity but never used to extract information from data.
They are derived within the microcanonical ensemble
by considering fragments interacting 
only by Coulomb and excluded volume,
which corresponds to the freeze-out configuration.
Within this ensemble, the statistical weight of a configuration
$c$, defined by the mass, charge and internal excitation energy
of each of the constituting $M_c$ fragments,
can be written~\cite{Radut02,I79-Bor13} as
\begin{eqnarray}
\nonumber
W_c(A,Z,E^*,V) = \frac1{M_c!} \chi V^{M_c} \prod_{n=1}^{M_c}\left( 
\frac{\rho_n(\epsilon_n)}{h^3}(mA_n)^{3/2}\right)
\\ 
\times
~ \frac{2\pi}{\Gamma(3/2(M_c-d))} ~ \frac{1}{\sqrt{({\rm det} I})}
~ \frac{(2 \pi K)^{(3/2)(M_c-d)-1}}{(mA)^{3/2}},
\label{eq:wc}
\end{eqnarray}
where $A$, $Z$, $E^*$ and $V$ are respectively the mass number,
the atomic number, the thermal excitation energy and
the freeze-out volume of the system.
$E^*$ is used up in fragment formation, fragment internal
excitation, fragment-fragment Coulomb interaction and
thermal kinetic energy $K$.
$I$ is the inertial tensor of the system whereas
$\chi V^{M_c}$ stands for the free volume or, equivalently, accounts for
inter-fragment interaction in the hard-core idealization.
$\rho_n(\epsilon_n)$ represents the fragment level density
(see ~\cite{Radut00} for the complete expression).
$d$ takes respectively values 0, 1 and 2 for energy, energy plus 
linear momentum and energy plus linear and angular momenta conservations.
In the following $d$ will be fixed to 1 to ensure coherence with
simulations which are used.
Taking into account that $S=\ln Z=\ln \sum_c W_c$, 
the microcanonical temperature is deduced from its statistical
definition~\cite{Radut02}:
\begin{eqnarray}
T=\left(\frac{\partial S}{\partial
E^*}\right)^{-1}&=&\left(\frac1{\sum_c W_c} \sum_c 
W_c(3/2M_c-5/2)/K\right)^{-1}\\
&=&\langle(3/2M_c-5/2)/K\rangle^{-1}.
\label{eq:T}
\end{eqnarray}
The notation $\langle\rangle$ refers to the average over the ensemble
states. The heat capacity of the system, $C$, is related to
the second derivative of the entropy by the equation
$\partial^{2}S /\partial E^{*2}$ = $-1/CT^{2}$.
Thus, one can evaluate the second derivative of the system entropy
versus $E$~(Eq.~(\ref{eq:15})) or alternatively the heat 
capacity $C$~(Eq.~(\ref{eq:14}))
\begin{eqnarray}
\nonumber
\frac{\partial^{2} S}{\partial
E^{*2}}&=&\langle \frac{(3/2M_c-5/2)(3/2M_c-7/2)}{K^2}\rangle
\\
&-&\langle\frac{(3/2M_c-5/2)}{K}\rangle^2
\label{eq:15}
\end{eqnarray}

\begin{eqnarray}
C=\left(1-T^2\langle \frac{(3/2M_c-5/2)(3/2M_c-7/2)}{K^2}\rangle\right)^{-1}
\label{eq:14}
\end{eqnarray}

These two quantities only depend on two parameters $M_c$ and $K$ which
must be estimated at freeze-out.

In Refs.~\cite{I58-Pia05,I66-Pia08} we presented simulations which
correctly reproduce most experimental observables and estimate
freeze-out properties in a fully consistent way.
They concern hot nuclei which undergo multifragmentation\\ 
formed in central 
collisions and selected by event shape sorting (quasifused systems, QF, from
$^{129}Xe$+$^{nat}Sn$, 32-50 AMeV)~\cite{I69-Bon08}. 
We will use the event by event properties
at freeze-out which were inferred to calculate the required 
quantities $M_c$ and $K$.

Experimental data were collected with the 4$\pi$
multidetector INDRA described in detail in Ref.~\cite{I3-Pou95,I5-Pou96}.
Accurate particle and fragment identifications were achieved
and the energy of the detected products was measured
with an accuracy of 4\%. Further details can be found 
in Ref.\cite{I14-Tab99,I33-Par02,I34-Par02}.

The method for reconstructing freeze-out
properties from simulations~\cite{I58-Pia05,I66-Pia08} 
requires data with a very high degree
of completeness, 
(measured fraction of the available charge $\geq$93\% 
of the total charge of the system),
crucial for a good estimate of Coulomb energy.
QF sources are reconstructed, event by event,
from all the fragments and twice the charged particles emitted in the range
$60-120^{\circ}$ in the reaction centre of mass, in order to exclude
the major part of pre-equilibrium
emission~\cite{I29-Fra01,I69-Bon08}; with such a prescription only particles
with isotropic angular distributions and constant average kinetic energies are
considered. In simulations,
dressed excited fragments and particles at
freeze-out are described by spheres at normal density.
Then the excited fragments subsequently deexcite while flying apart.
All the available asymptotic experimental information (charged
particle spectra, average and standard deviation of fragment velocity
spectra and calorimetry) is used \textit{to constrain the four free parameters
of simulations} to recover the data at each incident energy: \textit{the percentage
of measured particles which were
evaporated from primary fragments, the collective radial energy, a minimum
distance between the surfaces of products at freeze-out and a limiting
temperature for excited fragments}. All the details of simulations can be found
in Ref.~\cite{I58-Pia05,I66-Pia08}.
The limiting temperature,
related to the vanishing of level density for fragments~\cite{Koo87},
was mandatory to reproduce the observed widths of fragment velocity
spectra. With a single temperature (internal and kinetic temperatures
equal) the sum of Coulomb repulsion, collective energy, 
thermal kinetic energy directed at random
and spreading due to fragment
decays accounts only for about 60-70\% of those widths.
By introducing a limiting temperature, 
which corresponds to intrinsic temperatures for fragments
in the range 4-7 MeV (see figure 1 of~\cite{I79-Bor13}), 
the thermal kinetic
energy increases, due to energy conservation, thus producing the missing
percentage for the widths of final velocity distributions.
The agreement between experimental and simulated velocity spectra for
fragments, for the
different beam energies, is quite remarkable (see figure 3
of~\cite{I66-Pia08}). 
Relative velocities between fragment pairs were also compared
through reduced relative velocity correlation 
functions~\cite{Kim92,I57-Tab05}
(see figure 4 of~\cite{I66-Pia08}).
Again a good agreement is obtained between experimental data and
simulations, which
indicates that the retained method (freeze-out topology built up
at random) and the deduced parameters are sufficiently relevant
to correctly describe the freeze-out configurations, including volumes.

For hot nuclei produced in central collisions the total excitation energy,
$E_{tot}$, differs from the thermal one due to the presence of a radial collective 
expansion energy, $E_R$, and $E_{tot}$ = $E^*$ + $E_R$. 
To derive event by event at freeze-out $E_{tot}$, $E^*$, 
the multiplicity, $M_c$ ($M^{fo}$) and the total
kinetic energy, $K$, of the multifragmenting hot nuclei, the following equations
are used.
 
The limiting temperature for the fragments, $T_{lim}$, is 
introduced according to the
formalism of~\cite{Koo87}, which corresponds to the following definition 
for the intrinsic temperature of the fragments, $T_{frag}$:
\begin{equation}
\frac{1}{T_{frag}}=\frac{3}{2<K^{fo}>}+\frac{1}{T_{lim}}
\label{eq1}
\end{equation}
where $<K^{fo}>$ is the average kinetic energy of fragments and particles at
freeze-out. 
The equation used for calorimetry and to derive the sharing between
internal excitation energy and kinetic energy on the event by event
basis is the following:
\begin{eqnarray}
\lefteqn{\sum_{k=1}^{M_{cp}}K^k_{cp} + \Delta B_{cp}
+ M_n^{fo}<K^{fo}> + M_n^{evap}\theta_{frag} + \Delta B_n}
\nonumber\\
&=&(M^{fo}-1)<K^{fo}>
+\sum_{k=1}^{M_f}a_k\theta_{frag}^2
+\Delta B_{fo} + V_{Coul}^{fo}+E_R
\nonumber\\
&=&E_{tot} +\Delta B_{hn},
\label{eq2}
\end{eqnarray}
where $\theta_{frag}$ is equivalent to the temperature $T_{frag}$ in an
ensemble average.
$\Delta B$, $hn$, $K$, $M$, $cp$, $n$,
and $fo$ stand respectively for mass excess, hot nucleus, kinetic 
energy, multiplicities, charged products, neutrons and freeze-out.
\(E_R=\sum_{k=1}^{M^{fo}}({\frac{r_k}{R_0}})^{2}A_kE_0\) is the radial
collective expansion energy ($R_0$ is the rms of fragment and particle
distances to centre at the freeze-out volume, $E_0$ the radial expansion
energy at $R_0$ and $r$ is the distance of the considered
particle/fragment of mass A from the centre of the fragmented source).
$V_{Coul}^{fo}$ is the Coulomb energy of the configuration for
freeze-out previously determined. The chosen level
density parameter \(a_k=\frac{A_k}{10}MeV^{-1}\), where $A_k$ is the mass of
the $k^{th}$ fragment, well corresponds to that
expected for a typical primary fragment~\cite{Shl91}.
By introducing Eq.~(\ref{eq1}) in Eq.~(\ref{eq2}) we obtain 
a third degree equation in
$<K^{fo}>$, from which we can deduce the energy sharing between the internal
excitation energy of the fragments and the total thermal kinetic energy
at freeze-out $K = (M^{fo}-1)<K^{fo}>$. $K$ is shared at random
between all the particles and fragments at freeze-out under
constraints of conservation laws. Information on initial angular momenta of 
multifragmenting QF hot nuclei are unknown, which explains why $d$=1
is used for Eqs.~(\ref{eq:15}) and ~(\ref{eq:14}). 
Systematic error on $E^*$ was estimated around 1~AMeV.
\begin{figure*}
\centering
\resizebox{0.75\textwidth}{!}{
\includegraphics{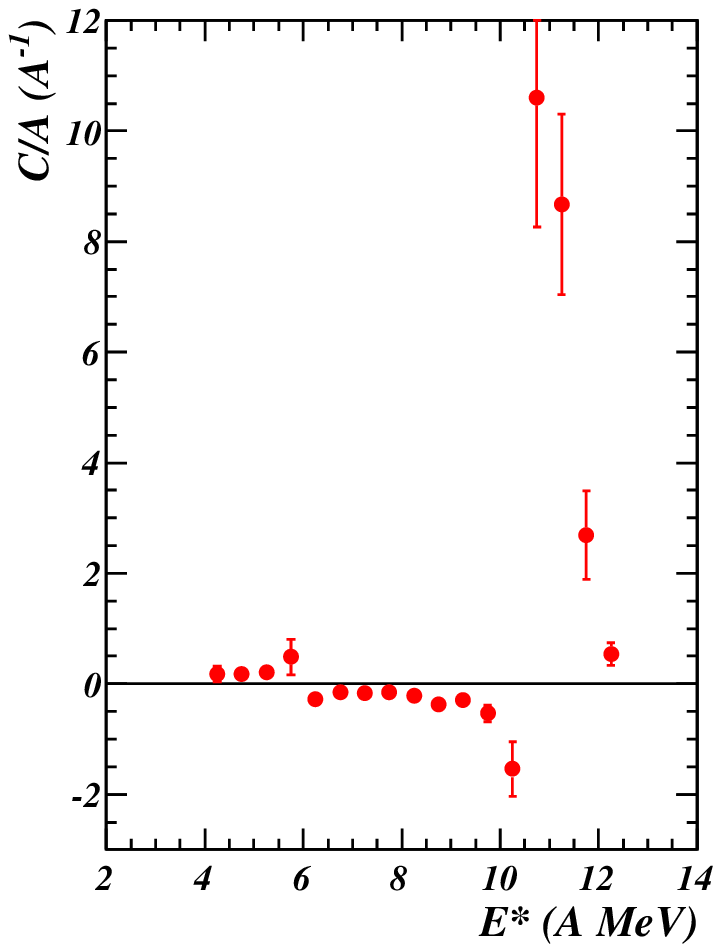}
\includegraphics{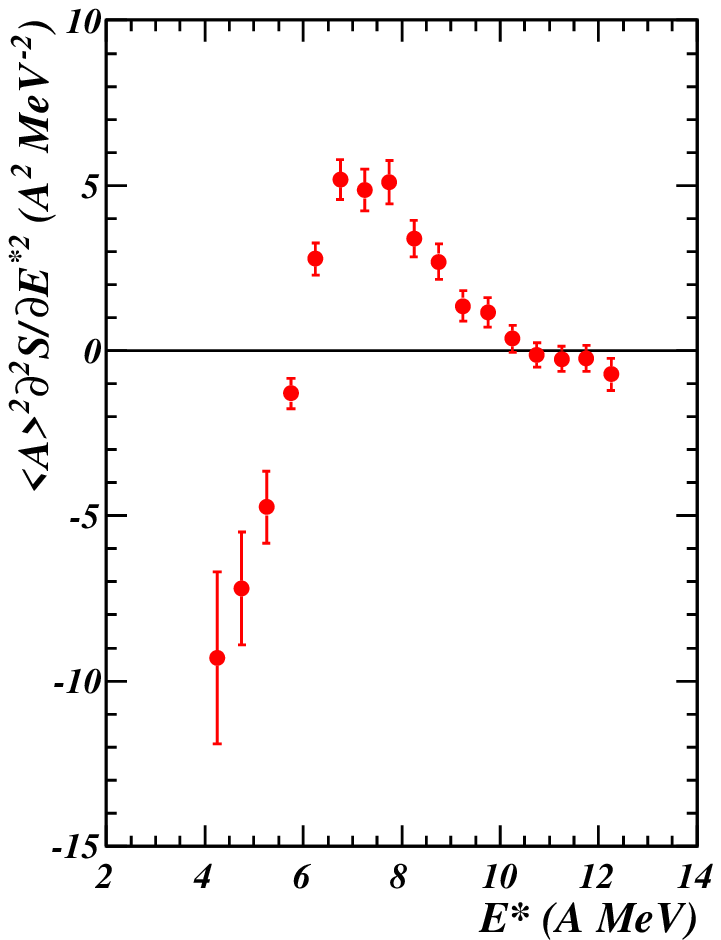}
}
\caption{ Heat capacity (left) and second derivative of the 
entropy (right) versus thermal excitation energy $E^*$. 
Error bars correspond to systematic plus statistical errors (see text).}
\label{fig:cneg}
\end{figure*}

Values of heat capacity and second derivative of the entropy versus thermal 
excitation energy $E^*$
have been calculated respectively from Eqs.~(\ref{eq:14}) and~(\ref{eq:15}) for QF 
hot nuclei
with $Z$ restricted to the range 80-100 to suppress tails of the distributions
at different beam energies. 
The average over the ensemble states have been assimilated to an
average over ``event ensembles'' sorted into $E^*$ bins.
A binning of 0.5 AMeV was chosen to have a sufficient number
of events in each bin in order to reduce statistical errors.
Figure~\ref{fig:cneg} shows the results.
Error bars correspond to systematic plus statistical
errors; systematic errors were evaluated by varying the free parameters
of simulations within their limits defined by a $\chi^2$ procedure~\cite{I66-Pia08}.
The left part of the figure shows the results for the direct calculation 
of $C/A$. Negative heat capacity is observed on a
rather large thermal excitation energy range and
the second diverging region is more visible than the first one.
\begin{figure}
\centering
\resizebox{0.35\textwidth}{!}{
\includegraphics{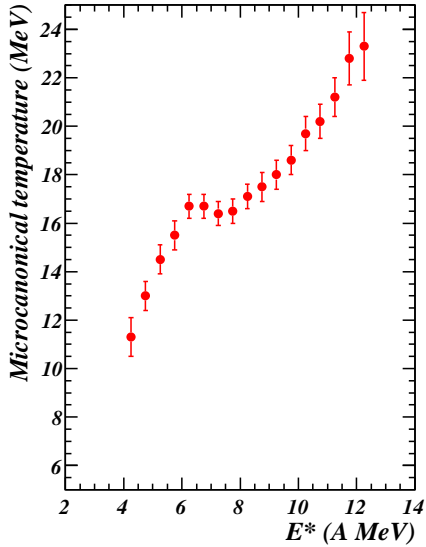}
}
\caption{Microcanonical temperatures calculated with Eq.~(\ref{eq:T}).
Error bars include systematic plus statistical errors.}
\label{fig:Tmicro}
\end{figure}
\begin{figure}
\centering
\resizebox{0.35\textwidth}{!}{
\includegraphics{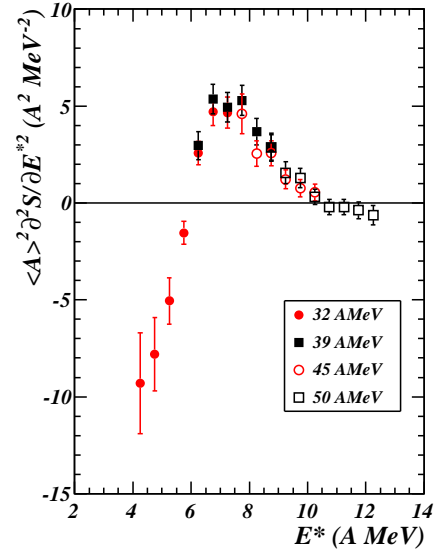}
}
\caption{Similar to the right part of fig.~\ref{fig:cneg}. The main contributions 
of the different bombarding energies are shown.}
\label{fig:cnegE}
\end{figure}
As the second derivative of the entropy is a very small 
quantity (from $\sim 2.10^{-4}$ to $\sim 3.10^{-6}$), 
we have kept the presentation made in~\cite{Radut02} i.e.
$A^{2} \partial^{2}S /\partial E^{*2}$ for fig.~\ref{fig:cneg} - right
part; A is replaced by the average mass, $<A>$, on the considered
$E^*$ bin. This quantity, which has small error bars in regions with
values close to zero, better defines the $E^*$ domain of negative heat capacity.  
Positive values are measured in the range 6.0 - 10.0 AMeV.
The related microcanonical temperatures calculated with 
Eq.~(\ref{eq:T}) are displayed in fig.~\ref{fig:Tmicro}.
Considering the error bars, they are rather constant around
17 - 18 MeV in the $E^*$ range where negative heat capacity 
is observed. With the large multiplicities observed in the present 
study, microcanonical temperatures are close to classical kinetic temperatures
(see figure 3 of~\cite{I79-Bor13}).
Last point, the statistical nature of the carefully selected event samples, i.e.
the independence of the entrance channel, was verified
and fig.~\ref{fig:cnegE} shows the main contributions of the different 
bombarding energies to the signal for each $E^*$ bin.

Negative heat capacity is thus confirmed for systems with $A$ around 200
without any constraint, which demonstrates once again the robustness
of this signal and supports the predictions made in~\cite{Cho00} i.e.
observation at constant pressure or at \textit{constant average volume.
It is important to stress here that negative heat capacity cannot be
observed at constant volume}. 
For the liquid-gas transition in hot nuclei
the volume is not fixed but multiplicity and partition-dependent, which
means that on the theoretical side one is forced to consider a
statistical ensemble for which the volume can fluctuate from event to
event around an average value~\cite{Bor19}.
\begin{figure}
\centering
\resizebox{0.35\textwidth}{!}{
\includegraphics{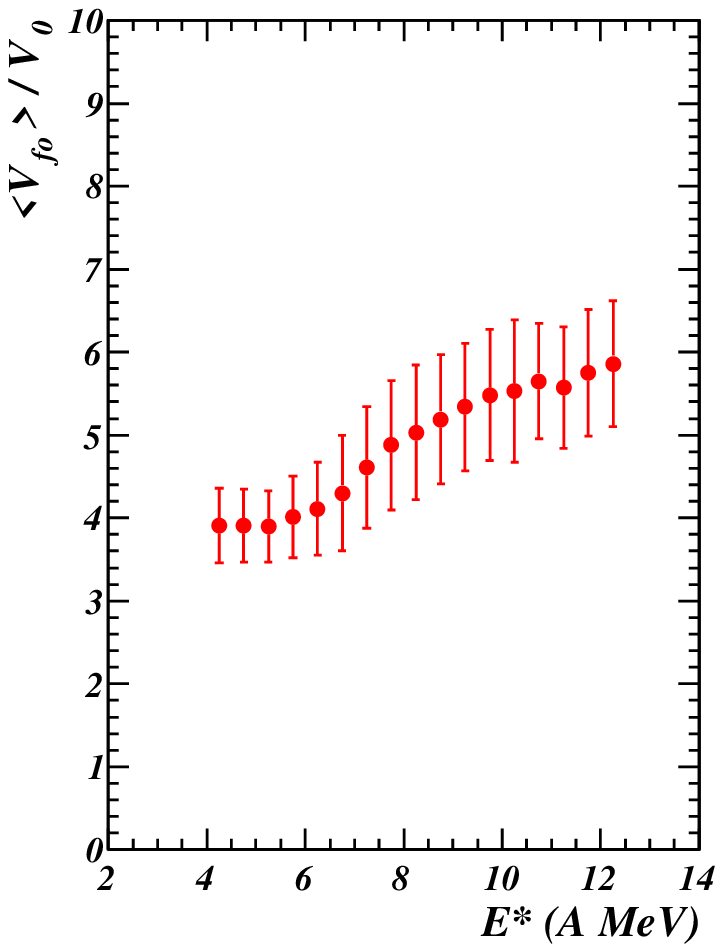}
}
\caption{Evolution of the average freeze-out volume of hot
nuclei, $V_{fo}$, with thermal excitation energy. Freeze-out volumes are normalized 
to volumes at saturation density, $V_0$, and error bars correspond to standard
deviations of $V_{fo}$ distributions.}
\label{fig:Vfo}
\end{figure}
\begin{figure}
\centering
\resizebox{0.35\textwidth}{!}{
\includegraphics{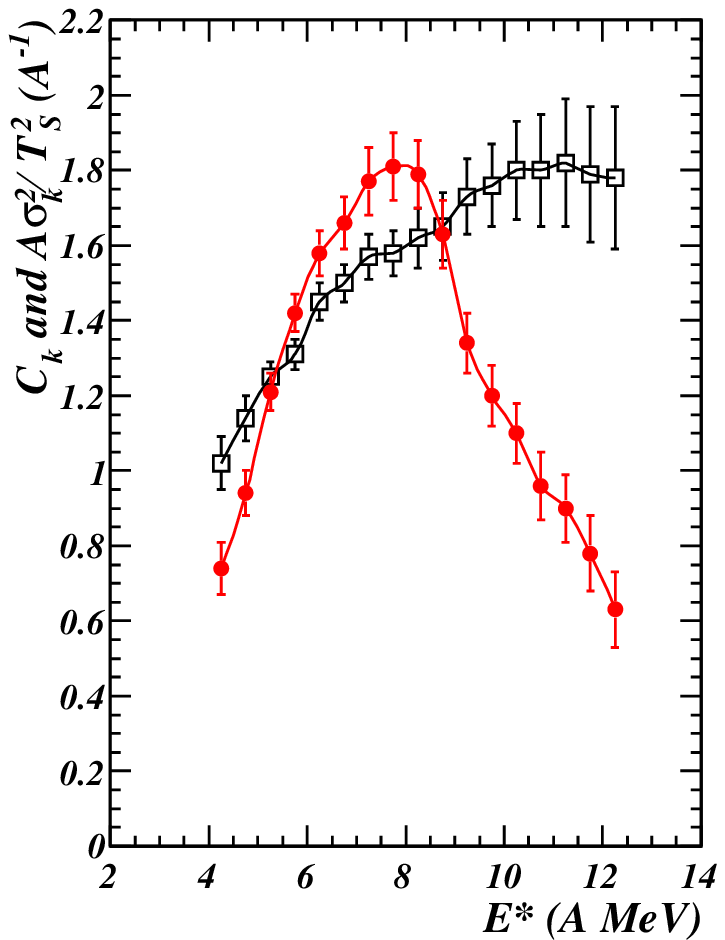}
}
\caption{Normalized kinetic energy fluctuations (filled circles) and
estimated $C_k$ values (open squares) related to the approximate
method (see text). Error bars correspond to systematic plus statistical errors.}
\label{fig:Cnegapp}
\end{figure}

As compared to previous estimates of heat
capacity~\cite{NicPHD,I46-Bor02}, there is a significant difference in the
$E^*$ range for negative values:  6.0$\pm$1.0 - 10.0$\pm$1.0 AMeV in the 
present study and $<$4.0$\pm$1.0 - 6.0$\pm$1.0 AMeV in the previous one.
For quasifusion hot nuclei selection the same shape event sorting was
used. The degree of completeness was different (93\% here to be compared to
80\% before) but it does not affect significantly the thermal
excitation energy \textit{per nucleon}. 
The method to reconstruct
freeze-out properties was also different.
But the main difference seems to be  related to the average freeze-out volume.
In previous estimates the
average freeze-out volume was kept constant at 3 times the volume at
normal density (3 $V_0$) over the whole thermal excitation energy range
whereas in the present study it varies 
from 3.9 to 5.9 $V_0$ (see fig.~\ref{fig:Vfo}).
A direct consequence in the approximate method is an increase of
Coulomb energy and consequently a decrease and a distortion with
excitation energy of
$<E_k>$ obtained by subtracting the potential part (Coulomb energy +
total mass excess) to the thermal energy. 
To verify this reasoning the approximate method has been used.
From our simulation, for each $E^*$ bin, we have calculated $<E_k>$
(see Eq.~(\ref{eq:Keos})) and derived an apparent single temperature,
$T_S$, needed to build the normalized kinetic energy fluctuations,
$A\sigma_{k}^{2}$/$T_{S}^{2}$, 
to be compared to $C_k$ (see Eqs.~(\ref{eq:CnegM1}) and~(\ref{eq:CnegM3})). 
Figure~\ref{fig:Cnegapp} shows that heat capacity becomes negative in
the $E^*$ range 5.5$\pm$1.0 - 9.0$\pm$1.0 AMeV, i.e. when
$A\sigma_{k}^{2}$/$T_{S}^{2}$ overcomes $C_k$.
This clearly confirms that the main difference, as compared to previous
estimates, comes from different average freeze-out volumes.
We also note a small decrease of the $E^*$ domain as compared to the
present study, which possibly comes from the previous method.
Last information, the apparent single temperature, $T_S$, exhibits a
monotonic behaviour, varying from
5.2 to 9.7 MeV over the whole $E^*$ range and values increasing from 6.5
to 8 MeV on the domain of negative heat capacity.

In conclusion, heat capacity measurements have been revisited without
approximation and by ignoring the hypothesis of a single temperature
associated with both internal excitation and thermal motion.
For those measurements microcanonical formulae and data
reconstructed at freeze-out with the help of a simulation have been used.
Negative heat capacity has been confirmed for hot nuclei with $A$ around 200
in the coexistence region of the phase transition.



\end{document}